\documentclass[aps,prl,twocolumn]{revtex4-1}

\usepackage{url}
\usepackage{graphicx}
\usepackage{amsmath}

\begin{document}

\title[Cell cycle regulation in bacteria]{Simultaneous regulation of cell size and chromosome replication in bacteria}
\author{Po-Yi Ho}
\author{Ariel Amir}
\email{arielamir@seas.harvard.edu}
\affiliation{School of Engineering and Applied Sciences, Harvard University, Cambridge, MA, USA}

\begin{abstract}
Bacteria are able to maintain a narrow distribution of cell sizes by regulating the timing of cell divisions. In rich nutrient conditions, cells divide much faster than their chromosomes replicate. This implies that cells maintain multiple rounds of chromosome replication per cell division by regulating the timing of chromosome replications. Here, we show that both cell size and chromosome replication may be simultaneously regulated by the long-standing initiator accumulation strategy. The strategy proposes that initiators are produced in proportion to the volume increase and is accumulated at each origin of replication, and chromosome replication is initiated when a critical amount per origin has accumulated. We show that this model maps to the incremental model of size control, which was previously shown to reproduce experimentally observed correlations between various events in the cell cycle and explains the exponential dependence of cell size on the growth rate of the cell. Furthermore, we show that this model also leads to the efficient regulation of the timing of initiation and the number of origins consistent with existing experimental results.
\end{abstract}

\maketitle


\section{Introduction}
Bacterial cells are extremely proficient in regulating and coordinating the different processes of the cell cycle. The Cooper-Helmstetter model proposes a molecular mechanism that couples two such processes, the replication of the chromosome and the division of the cell \citep{cooperhelmstetter}. In the model, cell division occurs a constant duration after the initiation of chromosome replication. The model implies a tight coordination between replication initiation and cell division such that in cells able to double faster than their chromosomes can replicate, multiple rounds of replications proceed simultaneously \citep{yoshikawa, cooperhelmstetter}. To answer how cells regulate the timing of initiation, it was proposed that “replication initiation factors” accumulate to a critical amount per origin of replication to trigger the initiation of replication \citep{baclifeseq}. Since the conception of the above model, many experiments and models have attempted to capture the molecular mechanisms responsible for the initiation of multiple rounds of replication. However, no model has been completely satisfactory \citep{blakely}.

As a result of the coupling between replication and division, the average cell size per origin is approximately a constant independent of the growth rate of the cell \citep{donachie68}. Furthermore, it is now understood that a common size regulation strategy for organisms including bacteria and budding yeast is the incremental model in which division occurs upon the addition of a constant size dependent on the growth rate of the cell \citep{AmirPRL, ilya, sattar, JW}. However, the molecular mechanisms responsible for the incremental model of size control remain in question.

Our main result in this work is to show that the initiator accumulation strategy not only regulates size according to the incremental model, but also regulates simultaneously the timing of initiation and the number of origins of replication. The strategy says that replication initiates upon the accumulation of a critical amount of replication initiation factors per origin. We emphasize the importance of the partitioning of replication initiation factors amongst origins, which we show is essential in order for the multiple replication forks to be adequately regulated. We show, analytically and numerically, that this strategy robustly regulates both cell size and the number of origins. Agreement between existing experiments and predictions of the above model reveals essential features that must be captured in any molecular mechanisms coordinating replication initiation and cell division. Finally, we make distinct predictions regarding the distribution of cell sizes at initiation of replication.


\section{Methods}
\subsection{Multiple origins accumulation model}

\begin{figure}
\begin{centering}
\includegraphics[width=3in]{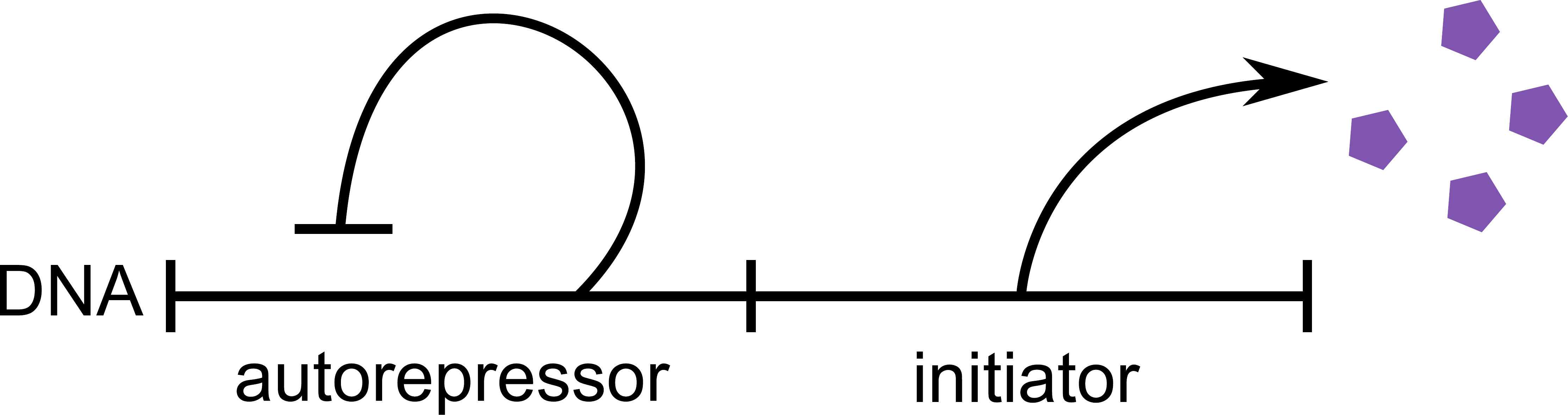}
\par\end{centering}
\caption{\label{fig:autorepressor} Schematic of the autorepressor model of initiator expression. An autorepressor is expressed such that its concentration remains constant and independent of the growth rate of the cell. The initiator is expressed in proportion to the autorepressor, but is localized at the origins of replication. Adapted from Sompayrac and Maaloe \citep{autorepressor}.}
\end{figure}

We consider the regulation strategy in which replication initiates upon the accumulation of a critical amount of replication initiation factors, or "initiators'', per origin of replication \citep{baclifeseq}. We assume that the initiators are expressed via an autorepressor model, as seen in Figure \ref{fig:autorepressor} \citep{autorepressor}. In this model, a protein is expressed such that its concentration $c$ remains constant and independent of the growth rate of the cell, which is plausible to achieve through autorepression. Therefore, an increase in the volume of the cell corresponds to a proportional increase in the copy number of this autorepressing protein. A second protein is the initiator and is expressed under the same promoter as the first, but in contrast to the first protein, it is \emph{localized} at the origins of replication. For simplicity, we assume that the initiators are equally partitioned amongst the origins. Initiation then occurs when a critical copy number per origin $N_{\textnormal{critical}}$ of the localized initiators is reached, after which the initiators are assumed to degrade. Under these assumptions, the copy number of the initiator effectively measures the increase in volume since initiation.

More precisely, if a cell initiated a round of replication at volume $v_{i}$ into $O$ number of origins, the amount of initiators $N_{\textnormal{initiators}}$ immediately after initiation is zero. To initiate the next round of replication, the cell must accumulate $ON_{\textnormal{critical}}$ initiators, but because the initiator is expressed under the same promoter as the autorepressor, the cell must also accumulate $ON_{\textnormal{critical}}$ autorepressors. Because the concentration of the autorepressor is constant, this implies that the cell must accumulate a corresponding volume $\Delta = N_{\textnormal{critical}}/c$ per origin, independent of the growth rate, to trigger the next initiation.

Thus, on a phenomenological level, the above biophysical model maps to the following regulation strategy for initiation,
\begin{equation}
v_{i}^{\textnormal{tot,next}}\approx v_{i}+O\Delta,\label{eq:repInitMod}
\end{equation}
Eq. \ref{eq:repInitMod} says that if a cell initiated a round of replication at cell volume $v_{i}$ into $O$ number of origins, then the cell will attempt to initiate another round of replication at total volume $v_{i}^{\textnormal{tot,next}}$, which is the sum of the volumes of all cells in the lineage since the initiation event at $v_{i}$ (typically two cells). This is not to be confused with the threshold model in which cells initiate upon reaching a threshold volume proportional to the number of origins, $v_{i}^{\textnormal{next}} \propto O$. For the rest of this manuscript, $O$ will denote the number of origins \emph{after} initiation at cell volume $v_{i}$ but \emph{before} initiation at total cell volume $v_{i}^{\textnormal{tot,next}}$.

\begin{figure}
\begin{centering}
\includegraphics[width=3in]{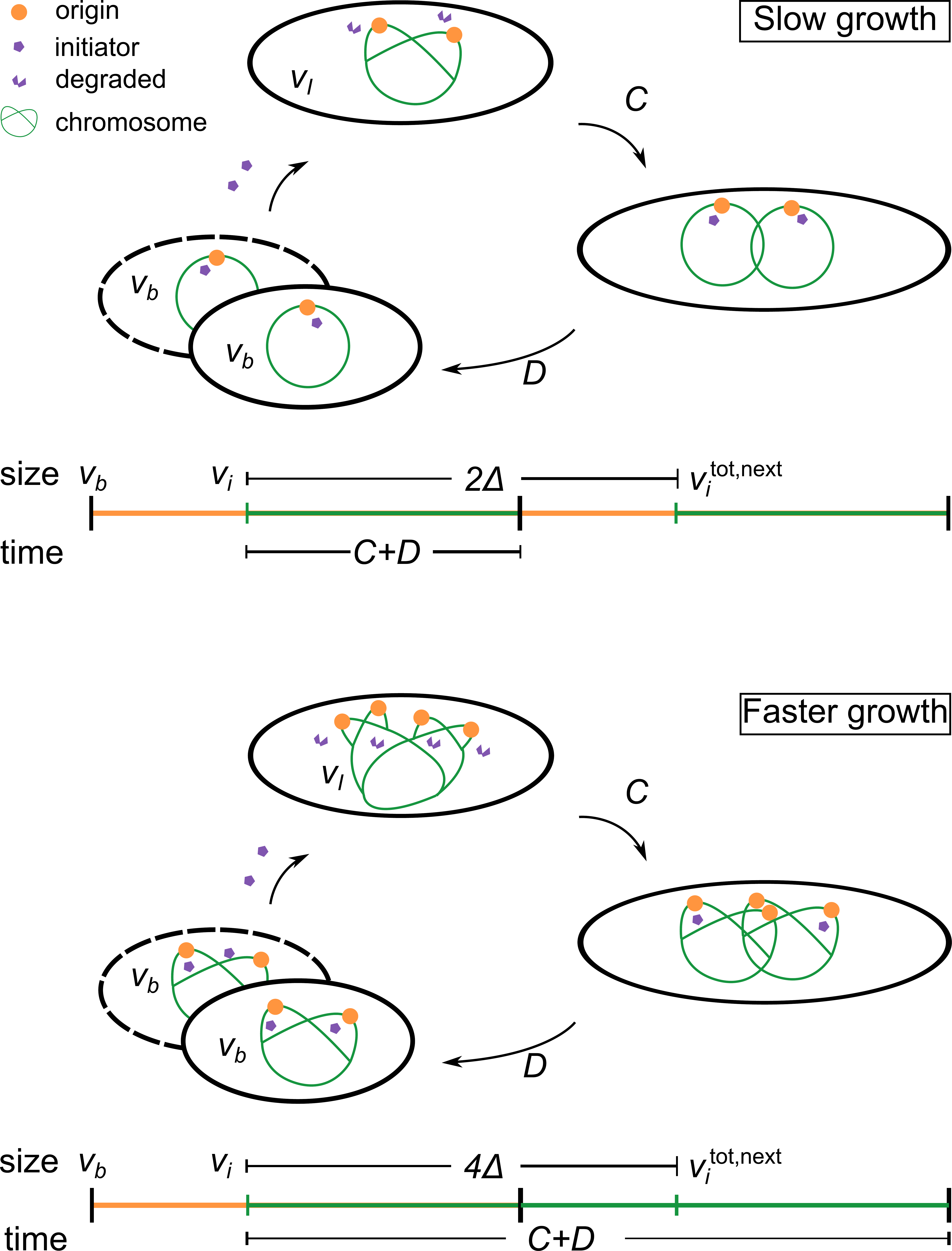}
\par\end{centering}
\caption{\label{fig:schematic} Schematic of the regulation strategy of the multiple origins accumulation model. See text for the details of the model. Slow growth denotes $0<\frac{C+D}{\tau}<1$. Faster growth denotes $1<\frac{C+D}{\tau}$. In the above example, $\frac{C+D}{\tau}<2$.}
\end{figure}

We assume an exponential mode of growth for cell volume with a constant doubling time $\tau$ and a corresponding constant growth rate $\lambda=\ln2/\tau$ \citep{manalis}. From Eq. \ref{eq:repInitMod} and the exponential mode of growth, durations between initiations are
\begin{equation}
t_{i}=\frac{1}{\lambda}\ln\left(1+\frac{O\Delta}{v_{i}}\right)+\xi,\label{eq:tinit}
\end{equation}
where $\xi$ represents some noise in the initiation process. An initiation event will trigger a division event after a constant duration $C+D$, where $C$ and $D$ are respectively the constant duration required to replicate the chromosome and the constant duration between replication termination and division \citep{cooperhelmstetter}. We will refer to Eq. \ref{eq:repInitMod} as the multiple origins accumulation model (i.e. initiators are accumulated per origin). Figure \ref{fig:schematic} illustrates this regulation strategy. We note that the strategy described here is mathematically equivalent to the "replisome" model of Bleecken \citep{bleecken} (not to be confused with the current use of the term replisome).

Finally, we will not take into account additional biological mechanisms that act at the level of the initiation of chromosome replication, such as \emph{oriC} sequestration, Dam methylation, and the "eclipse" phenomenon \citep{boganseq, kleckner, eclipse}. While these mechanisms are important to prevent rapid re-initiations, by themselves they are insufficient in ensuring an appropriately coordinated coupling between chromosome replication and cell division, which is the main focus of our work.

\subsection{Numerical simulations}

\begin{figure}
\begin{centering}
\includegraphics[width=3in]{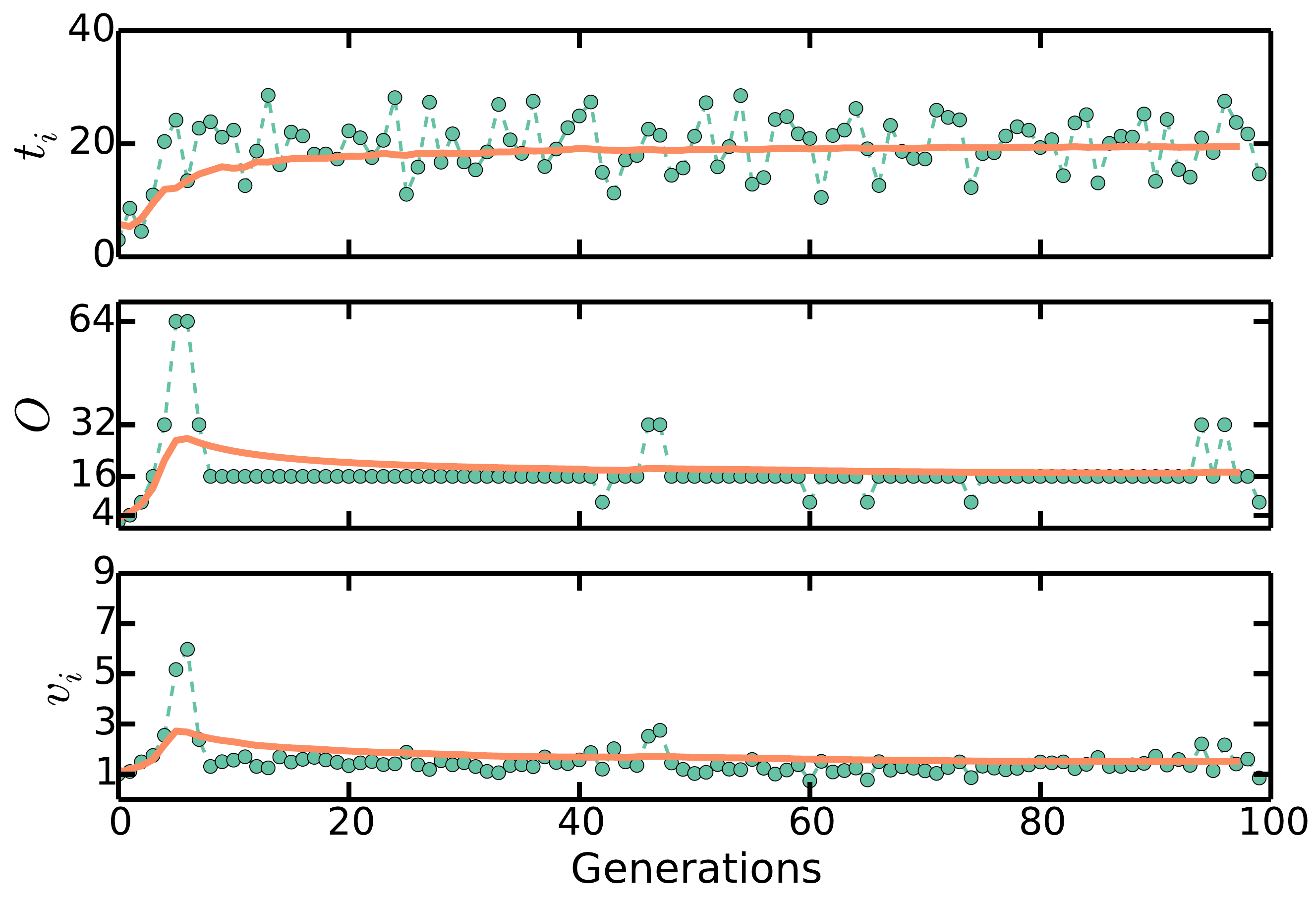}
\par\end{centering}
\caption{\label{fig:statdist} $t_{i}$, $O$, and $v_{b}$ approaching stationary distributions in numerical simulations of the multiple origins accumulation model. First, we initialize a population of $N$ cells with uniformly distributed cell ages. Durations between initiations of replication are calculated as Eq. \ref{eq:tinit} and the noise in the initiation process is assumed to be normally distributed. In an initiation event, the number of origins in a cell is doubled. The corresponding division event occurs after a constant time $C+D$. In a division event, the number of origins in a cell, along with the size of the cell, is halved, and two identical cells are generated. There are no division events without the corresponding initiation events. Following this procedure, a population of cells will robustly reach a stationary distribution of cell sizes and number of origins per cell regardless of initial conditions. The plots here track one lineage of cells. Here, $C+D=70\textnormal{ mins}$, $\tau=20\textnormal{ mins}$, and $\sigma_{\tau}=4\textnormal{ mins}$. These are biologically realistic choices. We set $\Delta=1/2^{(C+D)/\tau}$ so that $\left\langle v_{b}\right\rangle \approx1$.}
\end{figure}

\begin{figure}
\begin{centering}
\includegraphics[width=3in]{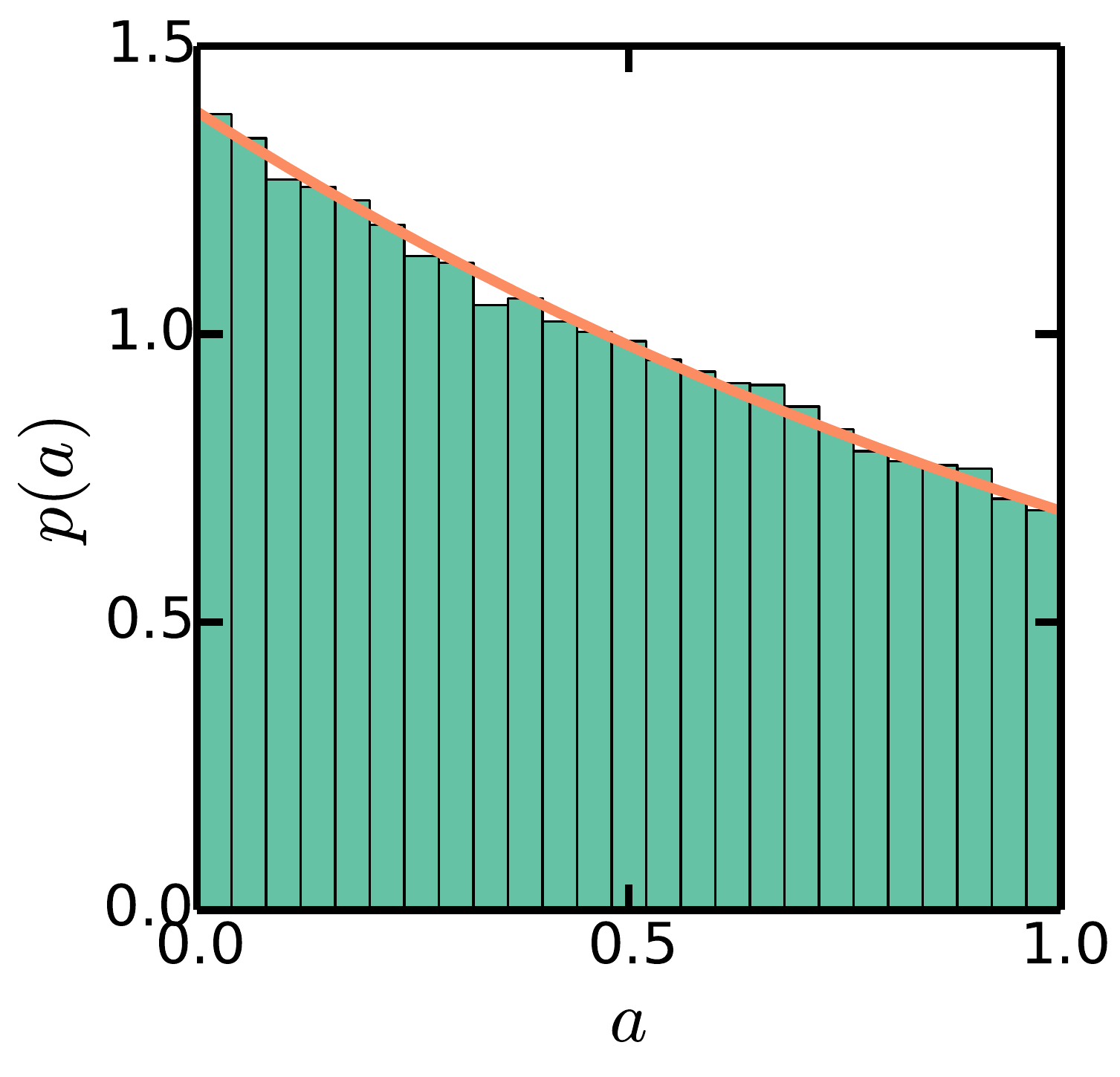}
\par\end{centering}
\caption{\label{fig:pastat} Stationary exponential distribution of cell ages. Simulations are the same as Figure \ref{fig:statdist}. The line plots $p\left(a\right)=\left(\ln2\right)2^{1-a}$, as described in the Appendix.}
\end{figure}

We can numerically simulate the multiple origins accumulation model given $C+D$, $\tau$, and $\Delta$ as experimentally measurable parameters. First, we initialize a population of $N$ cells with uniformly distributed cell ages. Durations between initiations are calculated as Eq. \ref{eq:tinit} and the noise in the initiation process is assumed to be normally distributed with standard deviation $\sigma_{\tau}$, though the precise nature of the noise does not affect any of our conclusions. It is assumed that in an initiation event, the number of origins in a cell is doubled. The corresponding division event occurs after a constant time $C+D$. In a division event, the number of origins in a cell, along with the size of the cell, is halved, and two identical cells are generated. We neglect the stochasticity arising from asymmetric divisions, which do not significantly affect any of the results. There are no division events without the corresponding initiation events. Following this procedure, a population of cells will robustly reach stationarity regardless of initial conditions, as seen in Figures \ref{fig:statdist} and \ref{fig:pastat}.

\section{Results}
\subsection{Multiple origins accumulation robustly and efficiently regulates the number of origins of replication}\label{sub:robust}

An important measurable consequence of the tight coupling between replication initiation and cell division is the average number of origins of replication per cell. It has been theoretically shown that the average number of origins per cell is
\begin{equation}
\left\langle O\right\rangle =2^{(C+D)/\tau}.\label{eq:I}
\end{equation}
The derivations leading up to Eq. \ref{eq:I}, summarized in Appendix, hinge on assuming an \emph{efficient process}: the population is growing exponentially and reaches a stationary distribution of cell ages, implying that there are no delays due to chromosome replication. These assumptions are independent of any molecular mechanisms for initiation and thus should be fulfilled by any efficient mechanism. We will show that the multiple origins accumulation model reproduces Eq. \ref{eq:I}.

The multiple origins accumulation model regulates initiation via negative feedback on the volume at initiation described by Eqs. \ref{eq:repInitMod} and \ref{eq:tinit}. The feedback enables cells to maintain a stationary average volume at initiation and a stationary average duration between initiations despite noise in the initiation process. Specifically, if a cell initiated replication at volume $v_{i}=O\Delta$, then the duration to the next initiation event is
\begin{equation}
t_{i}\left(O\right)\approx\frac{1}{\lambda}\ln\left(1+\frac{O\Delta}{O\Delta}\right) = \tau.\label{eq:tiniti}
\end{equation}
But if a cell initiated replication at a slightly larger volume $v_{i}=O\Delta+\delta v$, the duration to the next initiation event is
\begin{equation}
t'_{i}\left(O\right)\approx\frac{1}{\lambda}\ln\left(1+\frac{O\Delta}{O\Delta+\delta v}\right)\lesssim \tau.\label{eq:tinitis2}
\end{equation}
Eqs. \ref{eq:tiniti} and \ref{eq:tinitis2} say that a cell that initiated at a slightly larger volume than average tend to initiate again faster than average so that its volume at next initiation is again near the average. Similar reasoning says that cells that initiated at slightly smaller volumes tend to initiate again slower than average. In this way, cells maintain a stationary average volume at initiation and a stationary average duration between initiations.

Furthermore, the feedback enables cells to maintain a balanced cell cycle, in which there is on average one and only one initiation event per cell cycle. In the case of negligible noise, a balanced cell cycle implies that cells will initiate at cell age \citep{bremer96}
\begin{equation}
a_{i}=1+\lfloor\frac{C+D}{\tau}\rfloor-\frac{C+D}{\tau},\label{eq:ageinitiation}
\end{equation}
where $a=0$ represents cell birth, $a=1$ represents cell division, and $\lfloor\rfloor$ is the mathematical floor operator (largest integer smaller or equal to the argument). But in the case of realistic noise, a cell may initiate an extra round of replication if the noise is negative enough, $\xi/\tau\lesssim\lfloor (C+D)/\tau\rfloor-(C+D)/\tau$, which corresponds to an extra initiation at volume $v'_{i}=2O\Delta-\delta v$. The multiple origins accumulation model is robust to these stochastic events because a cell that initiated an extra round of replication will initiate again after
\begin{equation}
t_{i}\left(2O\right)\approx\frac{1}{\lambda}\ln\left(1+\frac{2O\Delta}{2O\Delta-\delta v}\right)\gtrsim \tau.\label{eq:tiniti2}
\end{equation}
In other words, cells with extra rounds of replication will initiate slower than those without so that the stationary average duration between initiations is maintained. The cell cycle following the extra initiation will typically not have any initiations, so that the initiation following the extra initiation will occur at approximately the average volume at initiation. A cell that missed a round of replication will return to a balanced initiation process in the analogous manner. In this way, the multiple origins accumulation model is able to efficiently maintain a balanced cell cycle in fast growth conditions. In contrast, the model simulated by Campos et al. \citep{JW} is not robust to the noise in the initiation process, because in their model, the incremental volume needed to trigger initiation is not partitioned between origins. We will elaborate on this in Section 3.3.

\begin{figure}
\begin{centering}
\includegraphics[width=3in]{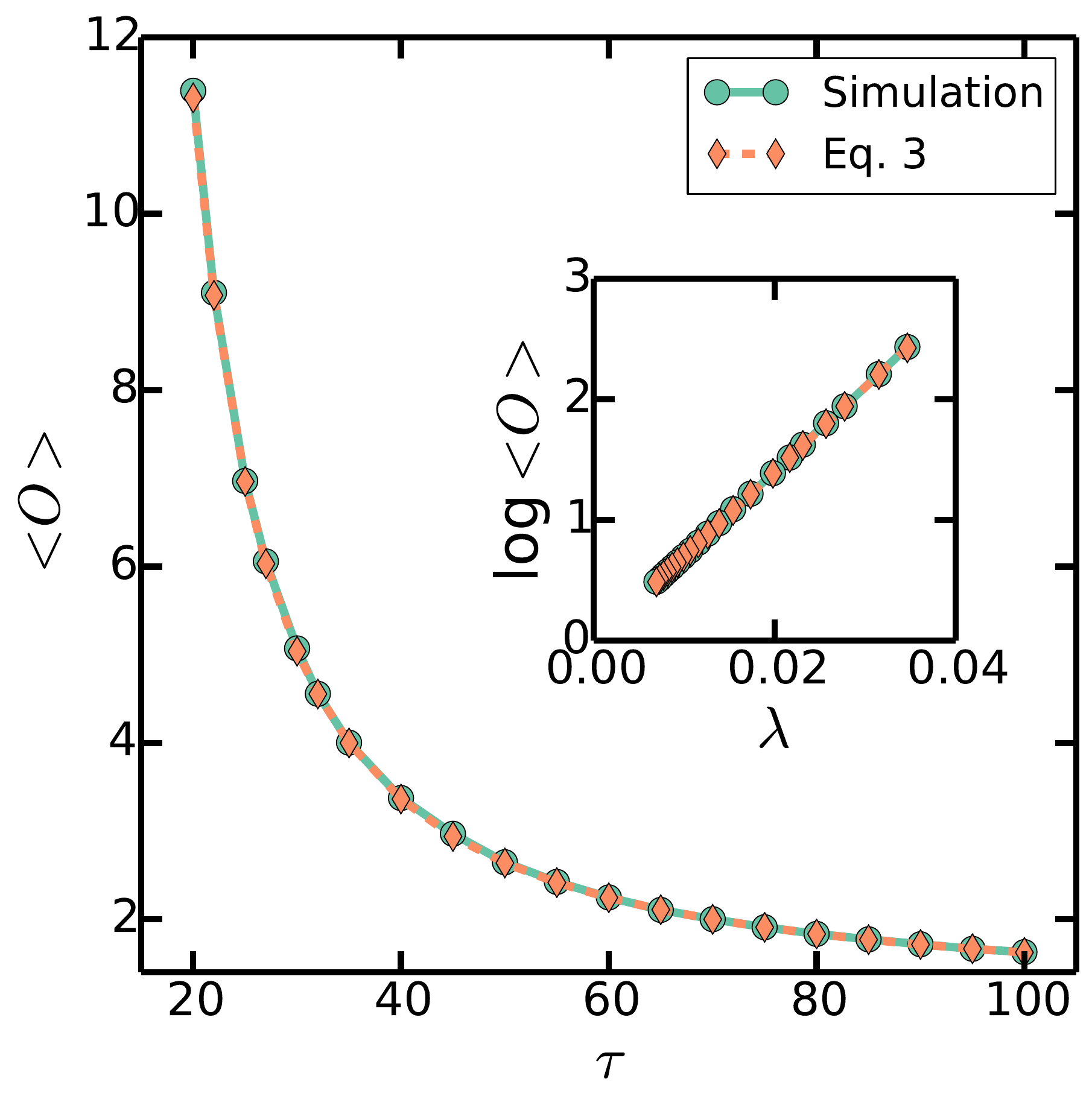}
\par\end{centering}
\caption{\label{fig:IvTd} $\left\langle O\right\rangle$ as a function of $\tau$ shows regulation of chromosome replication. Simulations are the same as Figure \ref{fig:statdist}, with a varying $\tau=20\textnormal{ to }100\textnormal{ mins}$, a fixed $C+D=70\textnormal{ mins}$, and $\sigma_{\tau}/\tau=0.2$. Dashed line plots Eq. \ref{eq:I}. Similarly, inset plots $\textnormal{log}\left\langle O \right\rangle$ as a function of $\lambda$}.
\end{figure}

The multiple origins accumulation model is therefore able to robustly regulate the timing of initiations in face of extra initiations. Extra initiations can occur not only because of noise in the initiation process, but also because of a shift in the growth rate of the cell such as that found in a shift-up experiment, in which a population of cells is abruptly switched from one nutrient condition to a richer nutrient condition allowing for faster growth. The increase in growth rate corresponds to a decrease in the duration between initiations. Therefore, cell in a shift-up experiment will initiate extra rounds of replication in the cycle immediately following the shift-up, but as we have seen, the multiple origins accumulation model is able to appropriately regulate the timing of initiations to reflect the new growth rate. Simulations of the multiple origins accumulation model reached stationary distributions of cell ages, durations between initiations, cell sizes, and number of origins per cell, regardless of initial conditions or the magnitude of the noise $\xi$ in Eq. \ref{eq:tinit}, as seen in Figures \ref{fig:statdist} and \ref{fig:pastat}. Simulations also show that the number of origins is regulated as in Eq. \ref{eq:I}, as seen in Figure \ref{fig:IvTd}. The above considerations show that the multiple origins accumulation model regulates the number of origins robustly and efficiently in face of noise in the initiation process.

\subsection{Multiple origins accumulation robustly regulates cell size}

It was recently shown that the multiple origins accumulation model of replication initiation reduces to the incremental model of size regulation \citep{AmirPRL}
\begin{equation}
v_{d}\approx v_{b}+v_{0}.\label{eq:vdvb}
\end{equation}
Eq. \ref{eq:vdvb} says that if a cell is born with volume $v_{b}$, then the cell will attempt to divide at volume $v_{d}$, where $v_{0}$ is the constant incremental volume from birth to division. In fact, $v_{0}$ can be expressed in terms of known parameters. First, if a cell initiated replication at volume $v_{i}$, then the corresponding division event will occur at total volume $v_{d}^{\textnormal{tot}}=v_{i}2^{(C+D)/\tau}$. But there will have been $\log_{2}O$ division events since initiation at $v_{i}$, so that the corresponding volume at birth is
\begin{equation}
v_{b}=\frac{v_{d}^{\textnormal{tot}}}{O}=\frac{v_{i}2^{(C+D)/\tau}}{O}.\label{eq:vbi}
\end{equation}
In a balanced cell cycle, the next initiation event will occur at total volume $v_{i}^{\textnormal{tot,next}}\approx v_{i}+O\Delta$ and the corresponding division event will occur at total volume $v_{d}^{\textnormal{tot},\textnormal{next}}=v_{i}^{\textnormal{tot,next}}2^{(C+D)/\tau}$. Similarly, there will have been $\log_{2}O$ division events since initiation at $v_{i}^{\textnormal{next}}$, so that the corresponding volume at division is
\begin{equation}
v_{d}=\frac{v_{d}^{\textnormal{tot},\textnormal{next}}}{O}=\frac{v_{i}2^{(C+D)/\tau}}{O}+\Delta2^{(C+D)/\tau}.\label{eq:vdinext}
\end{equation}
Therefore
\begin{equation}
v_{d}-v_{b}\approx\Delta2^{(C+D)/\tau}.\label{eq:v0}
\end{equation}
Within the multiple origins accumulation model, this derivation is valid for any $C+D$ and $\tau$ \citep{AmirPRL}.

The incremental model of size regulation predicts distributions, correlations, correlation coefficients, and scalings consistent with existing measurements \citep{AmirPRL, ilya, sattar, JW}. In particular, the average cell volume at birth
\begin{equation}
\left\langle v_{b}\right\rangle \approx\Delta2^{(C+D)/\tau}.\label{eq:vb}
\end{equation}
Eq. \ref{eq:vb} says that the average cell volume at birth is exponentially dependent on the growth rate, a well-known and well-tested result for \emph{E. coli} and \emph{B. subtilis} \citep{schaechter, sattar, bsub}. Simulations of the multiple origins accumulation model also confirm this result, as seen in Figure \ref{fig:Ivvb}. Thus, the multiple origins accumulation model robustly regulates cell size.


\subsection{Master accumulation predictions are inconsistent with existing experiments}\label{sub:alt}

Consider the regulation strategy
\begin{equation}
v_{i}^{\textnormal{tot,next}}\approx v_{i}+\Delta.\label{eq:repInitJW}
\end{equation}
Eq. \ref{eq:repInitJW} says that if a cell initiated a round of chromosome replication at cell volume $v_{i}$ with $O$ number of origins of replication, then the cell will attempt to initiate another round of replication at total volume $v_{i}^{\textnormal{tot,next}}$, where $\Delta$ is a constant volume independent of the growth rate. This is an incremental model of size control applied at initiation. If we assume the same mode of initiator expression as before, then the regulation strategy described by Eq. \ref{eq:repInitJW} corresponds to replication initiation upon accumulation of a critical amount of initiators \emph{without} partitioning of initiators between origins. Instead, a plausible molecular picture is that of initiators accumulating at a "master'' origin, whose initiation triggers the cascade initiation of other origins \citep{cascade}. We will therefore refer to Eq. \ref{eq:repInitJW} as the master accumulation model. In contrast to the multiple origins accumulation model described above, here in the master accumulation model, the total volume at next initiation does not depend on the number of origins present in the cell.

As before, we assume an exponential mode of growth with a constant doubling time $\tau$. The durations between initiations are therefore
\begin{equation}
t_{i}\left(O\right)=\frac{1}{\lambda}\ln\left(1+\frac{\Delta}{v_{i}}\right)+\xi,\label{eq:tinitJW}
\end{equation}
where $\xi$ represents some noise in the initiation process. Again as before, an initiation event will trigger a division event after a constant duration $C+D$. Eq. \ref{eq:tinitJW} differs from Eq. \ref{eq:tinit} by a missing factor of $O$. As we show below, the factor of $O$ is essential in regulating appropriately the timing of initiations and the master accumulation model does not reproduce the well-known exponential scaling of cell size with growth rate. The derivation follows and is similar to that of the multiple origins accumulation model. First, if a cell initiated replication at volume $v_{i}$, then the corresponding division event will occur at total volume $v_{d}^{\textnormal{tot}}=v_{i}2^{(C+D)/\tau}$. But there will have been $\log_{2}O$ division events since initiation at $v_{i}$, so that the corresponding volume at birth is
\begin{equation}
v_{b}=\frac{v_{d}^{\textnormal{tot}}}{O}=\frac{v_{i}2^{(C+D)/\tau}}{O}.\label{eq:vbiJW}
\end{equation}
In a balanced cell cycle, the next initiation event will occur at total volume $v_{i}^{\textnormal{tot,next}}\approx v_{i}+\Delta$ and the corresponding division event will occur at total volume $v_{d}^{\textnormal{tot},\textnormal{next}}=v_{i}^{\textnormal{tot,next}}2^{(C+D)/\tau}$. Similarly, there will have been $\log_{2}O$ division events since initiation at $v_{i}^{\textnormal{next}}$, so that the corresponding volume at division is
\begin{equation}
v_{d}=\frac{v_{d}^{\textnormal{tot},\textnormal{next}}}{O}=\frac{v_{i}2^{(C+D)/\tau}}{O}+\frac{\Delta2^{(C+D)/\tau}}{O}.\label{eq:vdinextJW}
\end{equation}
Therefore,
\begin{equation}
v_{d}-v_{b}=\frac{\Delta2^{(C+D)/\tau}}{O}.\label{eq:v0JW}
\end{equation}
This derivation is valid for any $C+D$ and $\tau$. However, from Eq. \ref{eq:I}, $O$ should scale exponentially with the growth rate like $2^{(C+D)/\tau}$ so that
\begin{equation}
\left\langle v_{b}\right\rangle \approx v_{d}-v_{b}\sim\Delta.\label{eq:vbJW}
\end{equation}
Eq. \ref{eq:vbJW} says that the average cell size is a constant roughly independent of the growth rate, a prediction contradicting the well-tested exponential scaling with growth rate for the model organisms mentioned above.

Furthermore, the above reasoning assumes that the master accumulation model can maintain a balanced cell cycle. But the master accumulation model cannot robustly maintain a balanced cell cycle in face of noise as Eq. \ref{eq:tinitJW} demonstrates. Specifically, if a cell initiated replication at volume $v_{i}=\Delta$, then the duration to the next initiation event is
\begin{equation}
t_{i}\left(O\right)\approx\frac{1}{\lambda}\ln\left(1+\frac{\Delta}{\Delta}\right) = \tau.\label{eq:tinitiJW}
\end{equation}
But the cell may proceed to initiate an extra round of replication if the noise is negative enough, $\xi/\tau\lesssim\lfloor (C+D)/\tau\rfloor-(C+D)/\tau$, which corresponds to an extra round of initiation at volume $v'_{i}=2\Delta-\delta v$. The next initiation will then occur after
\begin{equation}
t_{i}\left(2O\right)\approx\frac{1}{\lambda}\ln\left(1+\frac{\Delta}{2\Delta-\delta v}\right)\gtrsim\log_{2}\left(\frac{3}{2}\right)\tau.\label{eq:tiniti2JW}
\end{equation}
Implying that:
\begin{equation}
t_{i}\left(2O\right)\lesssim t_{i}\left(O\right)\label{eq:tinitineqJW}
\end{equation}
Eq. \ref{eq:tinitineqJW} says that cells with more origins will initiate faster than those with less, giving rise to cells with average durations between birth and division not equal to $\tau$. In other words, the master accumulation model does not robustly regulate the initiation process to maintain a balanced cell cycle. Indeed, simulations of the master accumulation model do not converge to a balanced cell cycle. Likewise, \cite{JW} carried out simulations of the master accumulation model and obtained "widely abnormal cell size distributions.'' Given the above inconsistent predictions, the master accumulation model can be ruled out as a possible regulation strategy for replication initiation.


\subsection{Multiple origins accumulation suggests that variations in $C+D$ are small}

\begin{figure}
\begin{centering}
\includegraphics[width=3in]{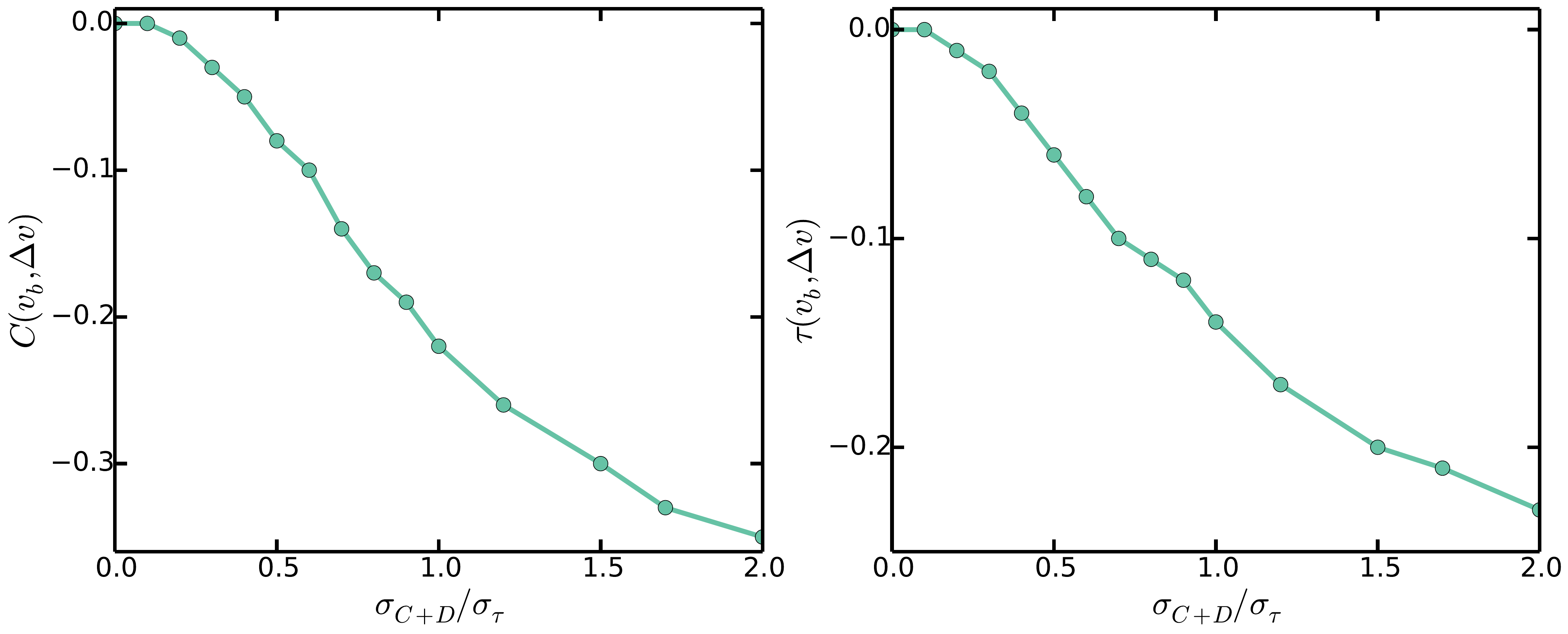}
\par\end{centering}

\caption{\label{fig:corrvsigr} Pearson (left) and Kendall (right) correlations between cell sizes at birth $v_b$ and cell size differences between birth and division $\Delta v$ against $\sigma_{C+D}/\sigma_{\tau}$ suggests that variations in $C+D$ are small.}
\end{figure}

In claiming that the master accumulation model gives incorrect correlations between growth rate dependent variables, \cite{JW} simulated the master accumulation model and reported negative correlations between cell size at birth $v_b$ and cell size differences between birth and division $\Delta v$, whereas none is observed experimentally. In contrast to claims in \cite{JW}, the negative correlations do not provide evidence against the multiple origins accumulation model nor the master accumulation model. Instead, the negative correlations provide evidence that variability in the durations from initiation to division $C+D$ should be small. Indeed, the multiple origins accumulation model, because of its reduction to the incremental model of size control, predicts no correlations between $v_b$ and $\Delta v$, \emph{given} that variations in $C+D$ are small compared to variations in $\tau$. Simulations assuming that durations from initiation to the corresponding division are normally distributed with mean $C+D$ and standard deviation $\sigma_{C+D}$ show that the correlations between $v_b$ and $\Delta v$ become increasingly negative as $\sigma_{C+D}/\sigma_{\tau}$ increases, as seen in Figure \ref{fig:corrvsigr}. Figure \ref{fig:corrvsigr} shows that as long as $\sigma_{C+D}/\sigma_{\tau} < 0.3$, the correlations between $v_b$ and $\Delta v$ will be close to zero. This is intuitive because when $\sigma_{C+D}$ is small compared to $\sigma_{\tau}$, fluctuations in cell sizes at birth arise due to variations in cell sizes at initiation, but these variations are negatively fed back into the multiple origins accumulation model as explained in Section \ref{sub:robust}. On the other hand, when $\sigma_{C+D}$ is comparable to $\sigma_{\tau}$, some fluctuations in cell sizes at birth arise due to variations in durations between initiation and division, but these variations are not accounted for by the multiple origins accumulation model. Variations of this nature give rise to the negative correlations between $v_b$ and $\Delta v$.

\subsection{Multiple origins accumulation predicts proportionality between cell size and the number of origins per cell}

\begin{figure}
\begin{centering}
\includegraphics[width=3in]{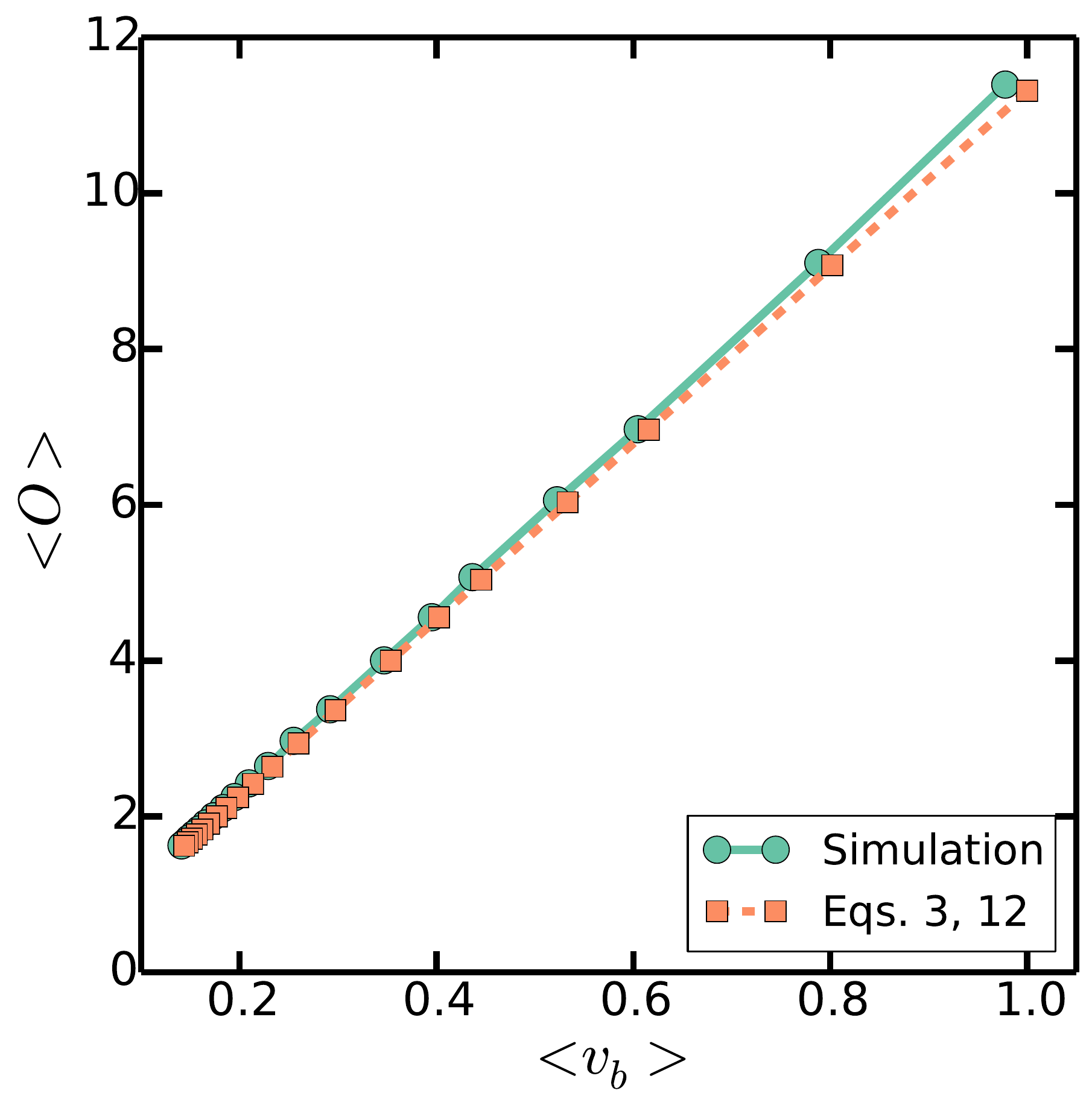}
\par\end{centering}
\caption{\label{fig:Ivvb} $\left\langle O\right\rangle$ against $\left\langle v_b\right\rangle$ shows proportionality between cell size and the number of origins per cell. Simulations are as in Figure \ref{fig:statdist}, with a varying $\tau=20\textnormal{ to }100\textnormal{ mins}$, a fixed $C+D=70\textnormal{ mins}$, and $\sigma_{\tau}/\tau=0.2$. Dashed line plots Eq. \ref{eq:I} and Eq. \ref{eq:vb}.}
\end{figure}

The simultaneous regulation of cell size and the number of origins per cell in the multiple origins accumulation model gives rise to a strict relationship between the two variables. In the multiple origins accumulation model, the average cell volume at birth Eq. \ref{eq:vb} is exponentially dependent on the growth rate, while the average number of origins per cell Eq. \ref{eq:I} also scales exponentially with the growth rate. Therefore we have that
\begin{equation}
\left\langle v_{b}\right\rangle \approx\Delta\left\langle O\right\rangle \label{eq:vbproptoi}
\end{equation}
That is, the multiple origins accumulation model predicts that given a fixed volume increment per origin $\Delta$, the average volume at birth and the average number of origins per cell will scale appropriately with respect to a varying $(C+D)/\tau$ to give rise to the above approximate proportionality. The critical size regulation strategy proposed by Donachie assumed this proportionality \citep{donachie68} , but is inconsistent with measured correlations in variables in \emph{E. coli} because thresholding size at any point in the cell cycle washes away the memory of the initial conditions, and therefore leads to a vanishing correlation coefficient between size at birth and size at division - contrary to measurements; on the other hand, multiple origins accumulation predicts this proportionality, and is consistent with measured correlations \citep{AmirPRL, ilya, sattar, JW}. Simulations of the multiple origins accumulation model confirm that cell size is indeed approximately proportional to the number of origins per cell, Figure \ref{fig:Ivvb}. We emphasize that the approximate proportionality is a property predicted by the multiple origins accumulation model. In contrast, other strategies that do not regulate the number of origins, such as the master accumulation model, would not predict it.

\subsection{Multiple origins accumulation predicts bimodal cell sizes at initiation}

\begin{figure}
\begin{centering}
\includegraphics[width=3in]{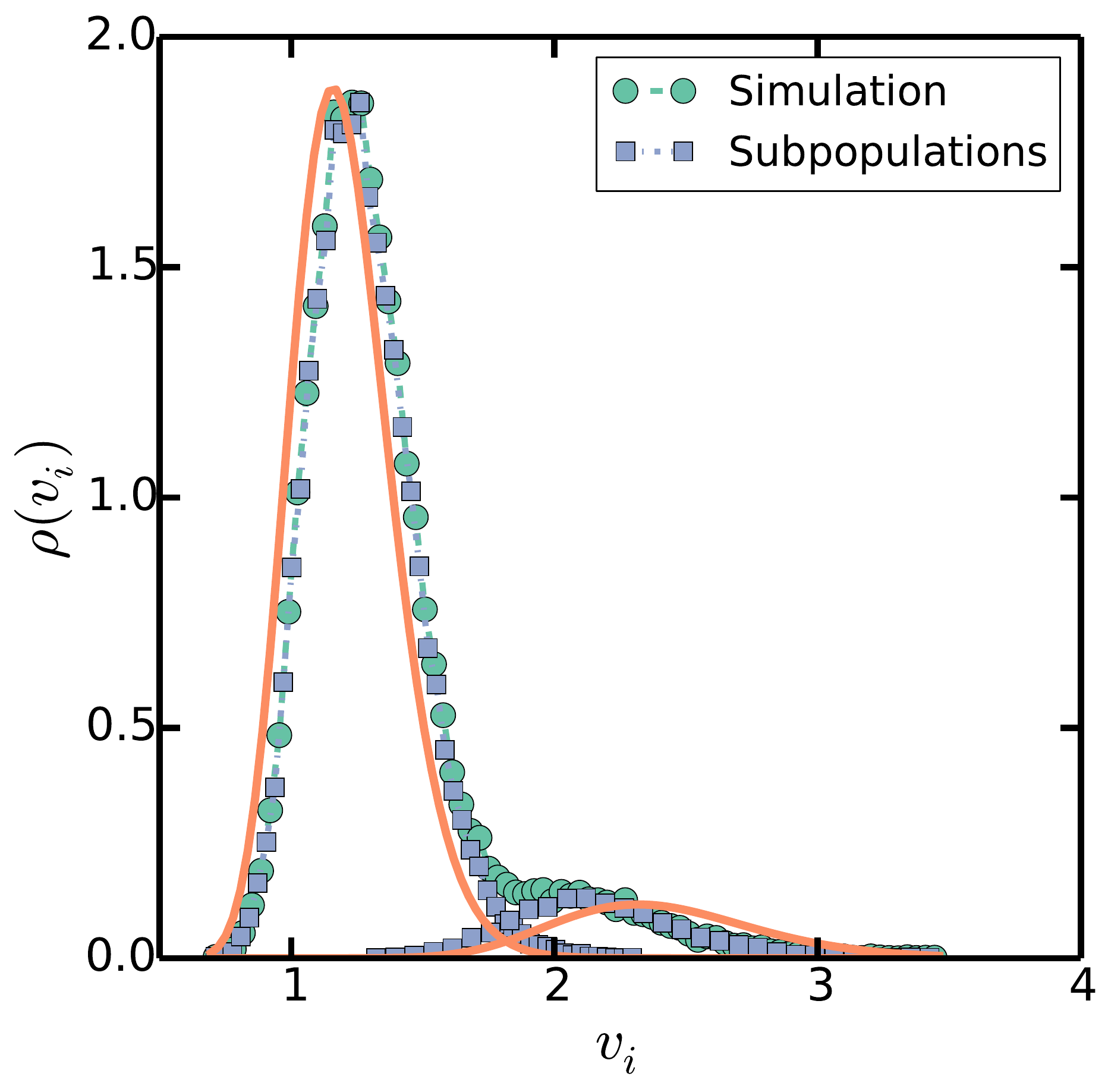}
\par\end{centering}
\caption{\label{fig:pvI} Distribution of volumes at initiation is bimodal. Simulations are as in Figure \ref{fig:statdist}, with $\tau=100\textnormal{ mins}$, $C+D=70\textnormal{ mins}$, and $\sigma_{\tau}=20\textnormal{ mins}$ as a specific, slow growth example. Square symbols separate volumes at initiation into two subpopulations, one whose cells initiated the predicted number of rounds of replication at volumes near $I_{0}\Delta$, and another whose cells initiated extra rounds of replication near twice that volume $2I_{0}\Delta$. Solid lines plot lognormal distributions with means $I_{0}\Delta$ and $2I_{0}\Delta$ and variances $\sigma_{v_{i}}^{2}=4\sigma_{\tau}^{2}/3\tau^{2}$ \citep{AmirPRL}.}
\end{figure}

In addition to the approximate proportionality between cell size and the number of origins per cell, the multiple origins accumulation model predicts that the distribution of cell sizes at initiation will be approximately bimodal because cells will initiate extra rounds of replication when the noise is large enough. Simulations show that the distribution of cell sizes at initiation is indeed bimodal, with one large peak corresponding to a subpopulation whose cells initiated the expected number of rounds of replication and a smaller subpopulation whose cells initiated extra rounds of replication, as seen in Figure \ref{fig:pvI}. Naively, because the distribution of cell sizes is lognormal in the multiple origins accumulation model, the distribution of cell sizes at initiation should be approximately the sum of two lognormal distributions with means $O_{0}\Delta$ and $2O_{0}\Delta$, where $O_{0}=2^{\lfloor (C+D)/\tau\rfloor}$, and the ratio between the frequencies of the two peaks equal to the probability that $\xi/\tau\lesssim\lfloor (C+D)/\tau\rfloor-(C+D)/\tau$. However, the value $O_{0}\Delta$ overestimates the average sizes at initiation of cells that initiated extra rounds of replication because of correlations between the volumes at initiation and the probability for extra rounds of replication. The correlations arise from Eq. \ref{eq:tinit}, which says that a smaller volume at initiation correlates with a larger probability for extra rounds of replication during the current cell cycle. The bimodal distribution of cell sizes at initiation highlights how the multiple origins accumulation model, without invoking other mechanisms, can robustly maintain a balanced cell cycle despite noise in the initiation process and is an experimentally testable prediction of our model.

\subsection{Multiple origins accumulation predictions are consistent with experiments on mutants}

Experiments on mutants of \emph{E. coli} and \emph{B. subtilis} produced results consistent with the predictions of the multiple origins accumulation model. Experiments have shown by manipulating the cell size of \emph{E. coli} via mutations that a decrease in $v$ is correlated with a decrease in $C+D$ and that a decrease in $v$ is also correlated with a decrease in $\left\langle O\right\rangle$ \citep{nhill}. In the language of the multiple origins accumulation model, $v$ is controlled by $\Delta$ and $(C+D)/\tau$. We assume that the mutations left unchanged the replication initiation mechanism in \emph{E. coli}, so that $\Delta$ is a constant throughout. But $\tau$ remained approximately constant with changes in $v$, so a decrease in $v$ must correspond to a decrease in $C+D$. One particular measurement reported for \emph{E. coli} cells an average doubling time $\tau\approx25\textnormal{ mins}$ for wildtype and mutant cells, $C+D\approx40\textnormal{ mins}+20\textnormal{ mins}$ for wildtype cells, and $(C+D)^{'}\approx30\textnormal{ mins}+20\textnormal{ mins}$ for mutant cells \citep{nhill}. Given these values and Eq. \ref{eq:vb}, the relative change in cell sizes corresponding to the reported difference in the durations from initiation to division should be $2^{\left((C+D)^{\prime}-(C+D)\right)/\tau}\approx0.75$. That is, the mutant cells should be $0.75$ times the size of the wildtype cells. This is in excellent agreement with the $25\%$ decrease in volume reported. Moreover, the size at initiation did not change for these smaller mutants, consistent with our model which predicts the size at initiation to depend only on $\Delta$ and not on $C+D$. In contrast, in the case of \emph{B. subtilis}, $C+D$ in smaller mutant cells remained constant, suggesting that $\Delta$ is the quantity which was changed. Based on this interpretation, our model would predict that both size at initiation and at birth would change proportionally. Indeed, it was found that both the average mutant cell size and the mutant cell size at initiation both decreased by approximately $35\%$ \citep{nhill}.  The above experiments also observed the predicted approximate proportionality between cell size and the number of origins per cell in both \emph{E. coli }and \emph{B. subtilis} \citep{nhill}. It remains to be shown that cell sizes at initiation fall into an approximate bimodal distribution. The agreement between experimental results and the predictions made by the multiple origins accumulation model speaks to the importance of regulating simultaneously cell size and the number of origins.


\section{Discussion}

The multiple origins accumulation model proposes that replication initiates upon the accumulation of a critical amount of initiators per origin. If the initiators are expressed as in the autorepressor model, this strategy corresponds to Eq. \ref{eq:repInitMod}, which in turn reduces to the incremental model of size control, which predicts distributions, correlations, and scalings consistent with existing measurements. Specifically, the average cell size scales exponentially with the growth rate of the cell, Eq. \ref{eq:vb}, as does the average number of origins per cell, Eq. \ref{eq:I}. The model robustly regulates both cell size and the number of origins per cell such that cell size is approximately proportional to the number of origins per cell, Eq. \ref{eq:vbproptoi}. These predictions are consistent with existing experiments on \emph{E. coli }and \emph{B. subtilis} \citep{nhill}. A proportionality between ploidy and cell size has also been observed in other organisms, including yeast \citep{what}. The multiple origins accumulation model is a general regulation strategy that may illuminate the source of the approximate proportionality between cell size and the number of origins across organisms.

An essential feature of the multiple origins accumulation model is the tight coupling between chromosome replication and cell division. The differences between the multiple origins accumulation model and the master accumulation model emphasizes this coupling and the importance of regulating the timing of initiation. By negatively regulating cell size in response to the number of origins via Eq. \ref{eq:repInitMod}, the multiple origins accumulation model is able to maintain a balanced cell cycle and achieve robustness in face of noise in the initiation process. However, the master accumulation model described by Eq. \ref{eq:repInitJW} is a regulation strategy without such a feedback mechanism. The master accumulation model is unable to maintain a balanced cell cycle and does not predict the exponential scaling of cell size. This suggests that regulation strategies must account for the number of origins per cell in order to regulate appropriately the frequency of division.

The coupling between the number of origins to the division frequency could be demonstrated via a shift-up experiment. It was found for \emph{E. coli} that cells maintain their rate of division for a duration of $C+D$ after a shift-up, a phenomenon known as rate maintenance \citep{kjeldgaard1958, shiftup}. The multiple origins accumulation model naturally accounts for rate maintenance, because division always occurs at time $C+D$ after initiation (chromosome replication rate is independent of growth rate). Furthermore, the model offers a robust mechanism to regulate, after a transient, the number of origins per cell appropriately with the new growth rate via Eq. \ref{eq:I}. The existence of a rate maintenance period implies that division is coupled to replication initiation, and the incremental model applied at birth and division is a valid phenomenological description only at stationarity. Instead, it is the underlying molecular mechanism of replication initiation that dictates the frequency of division.

Although the multiple origins accumulation model captures many aspects of the coupling between replication and division, experiments with minichromosomes suggest that the molecular mechanism is more complicated. Minichromosomes are plasmids containing the \emph{oriC} sequence coding for chromosomal origins. In general, minichromosomes initiate replications in coordination with chromosomes and do not affect the growth properties of the cell, such as the doubling time or the average cell size \citep{minichromexps}. However, if more than $\sim40$ minichromosomes are present in a cell, replication initiation is no longer synchronous, the doubling time increases, the average number of origins per cell decreases, and the average cell size decreases \citep{minichroms}. Another experiment inserted a second origin into \emph{E. coli} chromosome and observed again that the extra origin does not affect the growth properties of the cell \citep{twoori}. These result points to a more complicated molecular mechanism than accumulation of initiators per origin. Several mechanisms have been suggested, but none has been completely satisfactory. For example, the master accumulation mechanism discussed in Section \ref{sub:alt} is ruled out for being unable to robustly regulate the number of origins. Another plausible regulation strategy is one in which replication initiates when a critical ratio of active to inactive initiators is reached \citep{blakely}. The validity of this strategy remains to be tested.

The molecular mechanism underlying the regulation of replication initiation is yet to be unraveled, but here we have given significant constraints regarding the potential mechanisms. Specifically, this work and previous works have shown that the molecular mechanism in question should satisfy both the incremental model of size control and the mathematical form of the multiple origins accumulation model described by Eq. \ref{eq:repInitMod}, so that the predicted distributions, correlations, and scalings remain intact and consistent with existing experiments.


\section*{Acknowledgments}
The authors thank Lydia Robert and Nancy Kleckner for useful discussions.


%

\section*{Appendix: Derivations of the average number of origins per cell}

The average number of origins per cell has been calculated previously in two distinct derivations \citep{cooperhelmstetter, bremer}. The two derivations seemed to make different assumptions to arrive at the same conclusions, bringing into question the necessity of the underlying assumptions. Here, we reproduce the two derivations and show that both derivations in fact make the same assumptions.

The model of the cell cycle under consideration is due to Cooper and Helmstetter \citep{cooperhelmstetter}. In this model, replication initiation occurs on average every doubling time $\tau$. An initiation event then triggers a division event after a constant duration $C+D$, where $C$ and $D$ are respectively the constant duration required to replicate the chromosome and the constant duration between replication termination and division. Given $C$, $D$, and $\tau$, we want to find the average number of origins per cell. The average number of origins per cell is defined as $\left\langle O\right\rangle =\left\langle O_{\textnormal{total}}/N\right\rangle $, where $O_{\textnormal{total}}$ is the total number of origins in a population of cells, $N$ is the number of cells in that population, and brackets denote the ensemble average.

First, we reproduce the derivation due to Cooper and Helmstetter \citep{cooperhelmstetter}. To calculate the average number of origins per cell, we must first define the probability distribution underlying the ensemble average. In an asynchronous population of exponentially growing cells, the cells must be exponentially distributed in the cell cycle for ensemble averages to be stationary with respect to time \citep{powell}. Defining cell age $a=0$ at birth and $a=1$ at division, the exponential distribution of cell ages is
\begin{equation}
p\left(a\right)=\left(\ln2\right)2^{1-a}.\label{eq:pasupp}
\end{equation}
We can now calculate the desired ensemble averages. For example, if $0<\frac{C+D}{\tau}<1$, then a cell younger than $\left(\tau-(C+D)\right)/\tau$ will not be replicating its chromosome and will have only one origin, whereas a cell older than $\left(\tau-(C+D)\right)/\tau$ will be replicating its chromosome and will have two origins. We have assumed that the amount of time a cell spends with more than two origins is negligible, which is plausible for weak noise in the initiation process. The average number of origins per cell in an asynchronous, exponentially growing population is then
\begin{eqnarray}
\left\langle O\right\rangle  & = & \ln2\left[\left(\int_{0}^{\frac{\tau-(C+D)}{\tau}}2^{1-a}da\right)+2\left(\int_{\frac{\tau-(C+D)}{\tau}}^{1}2^{1-a}da\right)\right]\nonumber \\
 & = & \left[\left(2-2^{(C+D)/\tau}\right)+2\left(2^{(C+D)/\tau}-1\right)\right]\nonumber \\
 & = & 2^{(C+D)/\tau}.\label{eq:mIex}
\end{eqnarray}
Similarly if $1<\frac{C+D}{\tau}<2$, then a cell must initiate replication not only for its daughter cells, but also for its granddaughter cells. In this case, a cell with cell age less than $\left(2\tau-(C+D)\right)/\tau$ will not be replicating its chromosome for its granddaughters and will have only two origins, whereas a cell with cell age more than $\left(2\tau-(C+D)\right)/\tau$ will be replicating its chromosome for its granddaughters and will have four origins. Again, we have assumed weak noise.

Thus, we see that
\begin{align}
p\left(O=O_{0}\right) & =\ln2\int_{0}^{\Delta T}2^{1-a}da,\\
p\left(O=2O_{0}\right) & =\ln2\int_{\Delta T}^{1}2^{1-a}da,
\end{align}
where $O_{0}=2^{\lfloor (C+D)/\tau\rfloor}$, and $\Delta T=\left(\lfloor (C+D)/\tau\rfloor+1\right)\tau-(C+D)$. Simplification gives
\begin{equation}
\left\langle O\right\rangle =2^{(C+D)/\tau},\label{eq:mI}
\end{equation}
which generalizes Eq. \ref{eq:mIex} and is valid for any $C+D$ and $\tau$. The two assumptions made in this derivation are that the population is growing exponentially and that the population has reached a stationary distribution of cell ages.

Next, we reproduce the derivation due to Bremer and Churchward \citep{bremer}. Assuming exponential growth, the number of cells must grow exponentially as $N\propto2^{t/\tau}$. Similarly, the total number of origins must grow at the same exponential rate so that $O_{\textnormal{tot}}\propto2^{t/\tau}$. But an initiation event triggers a division event after a constant duration $C+D$, so the number of cells must on average lag behind the total number of origins by $2^{(C+D)/\tau}$. The average number of origins per cell must then be $\left\langle O\right\rangle =\left\langle O_{\textnormal{tot}}/N\right\rangle =2^{(C+D)/\tau}$. Although the distribution of cell ages was not explicitly involved in this derivation, the assumption of a stationary ensemble average under exponential growth is satisfied if and only if the distribution of cell ages is exponential \citep{powell}.

The exponential distribution of cell ages is not always realized in experimental setups. For example, single-cell experiments that track a lineage of cells, such as those in \cite{sattar}, will follow a different distribution, as discussed in \cite{lydiabmc}. Experiments that track a single cell will follow a uniformly distributed cell age. In that case, Eq. \ref{eq:mI} is replaced by
\begin{equation}
\left\langle O\right\rangle =O_{0}\left(1+\frac{C+D}{\tau}-\lfloor\frac{C+D}{\tau}\rfloor\right).\label{eq:mI2}
\end{equation}
Simulations tracking a population of cells with uniformly distributed cell ages confirm this result. The differences between Eq. \ref{eq:mI} and Eq. \ref{eq:mI2} do not significantly change the predictions of the multiple origins accumulation model.

\end{document}